# Generation of tightly compressed solitons with a tunable frequency shift in Raman-free fibers

Rodislav Driben[1,2,*] and Boris A. Malomed[2]

[1] Department of Physics & CeOPP, University of Paderborn, Warburger Str. 100, D-33098 Paderborn, Germany
[2] Department of Physical Electronics, Faculty of Engineering, Tel Aviv University, Tel Aviv 69978, Israel
*Corresponding author: driben@post.tau.ac.il

Optimization of the compression of input N-solitons into robust ultra-narrow fundamental solitons, with a tunable up- or downshifted frequency, is proposed in photonic crystal fibers (PCF) free of the Raman effect. Due to the absence of the Raman self-frequency shift, these fundamental solitons continue propagation maintaining the acquired frequency, once separated from the input N-soliton's temporal slot. A universal optimal value of the relative strength of the third-order-dispersion (TOD) is found, providing the strongest compression of the fundamental soliton, is found. It depends only the order of the injected N-soliton. The largest compression degree significantly exceeds the analytical prediction supplied by the Satsuma-Yajima (SY) formula. The mechanism behind this effect, which remains valid in the presence of the self-steepening, is explained.

OCIS Codes: 190.4370, 190.5530, 060.5295, 060.2630

In addition to fundamental solitons, optical media described by nearly-integrable equations [1] support higher-order N-solitons, with $N \geq 2$, which are oscillating modes periodically restoring their shape at distances that are multiples of the fundamental soliton period [2]. Observations of higher-order solitons in experimental settings were later reported [3]. Higher-order linear and nonlinear effects, such as the Raman-induced self-frequency shift (SFS) [4], self-steepening, and higher-order dispersion (HOD), described by the generalized nonlinear Schrödinger equation [5], cause fission of N-solitons [6-10]. The distance at which the fission occurs usually corresponds to the point where the bandwidth of the evolving N-soliton reaches its maximum. The fission of a higher-order solitons is a key mechanism for the creation of ultrashort frequency-tuned fundamental solitons [11-14] and generation of ultra-broadband optical supercontinuum [5, 8, 15].

In case of very weak perturbation, the initial N-soliton gives rise to N distinct solitons with analytically predicted peak powers and durations [2, 7]. On the other hand, in the case of strong HOD the fission is governed by the recently introduced Newton's cradle (NC) mechanism, which remains valid in the presence of the Raman and self-steepening effects [10]. It explains the ejection of fundamental quasi-solitons from the parental N-soliton temporal slot. In the case of strong HOD, along with discretely ejected solitons, the N-soliton fission gives rise to emission of strong dispersive radiation. Due to the interaction with the radiation, weaker solitons can acquire an additional acceleration, which leads to their collisions with earlier emitted stronger ones [16-21]. Thus the control over the generated fields deteriorates.

While launching picosecond pulses very close to the zero-dispersion wavelength of the fiber supporting the domination of the third-order dispersion (TOD) term, working with pulses shorter than 1 ps in the temporal domain requires consideration of the Raman and shock terms [3]. For longer ps pulses, instabilities may cause a break-up into a train of ultrashort fs soliton-like pulses. Recent advances in manufacturing PCFs filled with Raman-inactive gases [22, 23] provide an opportunity to deal with the Raman-free propagation, even for sub-picosecond pulses.

The present work aims to demonstrate a possibility to develop a source of tightly compressed stable solitons, with a frequency shift tunable over a significant wavelengths range. In the absence of the SFS and noise associated with this effect, fission-produced strongly compressed solitons, once generated, propagate keeping a constant internal frequency. We find a regime of optimal compression and frequency shift of the generated fundamental solitons, and demonstrate that the optimal value of the relative TOD strength depends solely on the order of the input N-soliton. The optimal values of the TOD strength are found to located between those corresponding to the fission of the initial N-soliton into N distinct fundamental solitons (the *Satsuma-Yajima (SY) regime* [2]), and the strongly pronounced NC regime [11] with a strong effect of radiation on ejected fundamental solitons. The mechanism of the generation of the frequency-shifted solitons is also valid for femtosecond pulses, actually being upheld by the action of the self-steepening term in the Raman-free fiber [22].

The propagation of ultrashort pulse's amplitude $u$ along the distance $z$ in the Raman-free fiber is governed by the extended nonlinear Schrödinger (NLS) equation in the normalized form [5, 22]

$$iu_z + (1/2)u_{TT} + |u|^2 u = i\delta_3 u_{TTT} - is(|u|^2 u)_T, \quad (1)$$

where $s = (\omega_0 T_0)^{-1}$ is the strength of the shock term, and $\delta_3 = \beta_3 / (6|\beta_2|T_0)$ is the normalized TOD coefficient, which will play the role of the control parameter in the analysis. We neglect the influence of the shock term at first ($s = 0$), to focus on the TOD effects. Then it will be shown that the shock term only enhances these effects. If one launches a fundamental soliton into the fiber with the TOD, the pulse will experience a small frequency shift of several nm, with a resonant dispersive wave generated at the front or rear side of the soliton, depending on the sign of $\beta_3$ [15]. The strength of the TOD depends on the proximity of the soliton to the zero-dispersion wavelength. Accordingly, the peak power of the pulse decreases with the propagated distance due to the emission of radiation. The situation is completely different if a higher-order $N$-soliton $u_0 = N\sqrt{P_0}\,\text{sech}(T/T_0)$ is launched into the fiber, where $T_0$ and $P_0$ are the width and peak power of the corresponding fundamental soliton. If $\beta_3$ is very small, it may be considered as a weak perturbation. In this case, the peak power $P_j$ of each fundamental soliton emerging from the fission of the $N$-soliton is

$$P_j = (2N - 2j + 1)^2 / N^2 \qquad (2)$$

[2]. For example, the 10-soliton will generate fundamental solitons with the largest peak power 3.6 times higher than that of the original $N$-soliton. Larger $\beta_3$ can produce a significant increase in the largest fundamental-soliton's peak power, along with the increase of its frequency shift. However, the further increase of $\beta_3$ leads to loss of the peak-power enhancement due to violent emission of radiation [10]. Thus, there should be an optimal value of $\beta_3$ for each $N$, which provides for the highest peak power of the lead soliton, and the corresponding frequency shift.

The results produced by simulations of Eq. (1) with $s = 0$ are displayed in Figs. 1,2 and summarized in Fig. 3. The blue curve in Fig. 1 demonstrates the fission of the 10-soliton in the case of a very small $\delta_3 = 0.002$. This result is compatible with Eq. (3). A larger value, $\delta_3 = 0.01$, provides a significant enhancement, as concerns the peak power and frequency shift of the lead output soliton (the black curve). However, further increase of $\delta_3$ leads to degradation of the power enhancement (the green and orange curves), due to the increasing radiation loss.

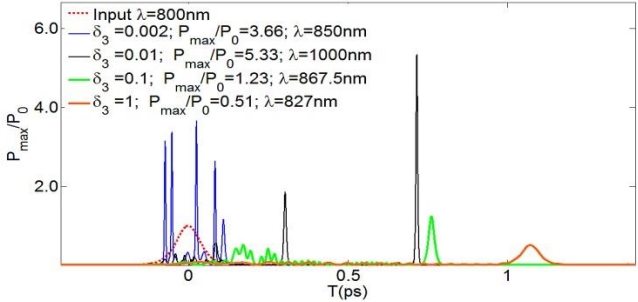

Fig. 1. (Color online) The generation of frequency-shifted fundamental solitons from the 10-soliton input with $T_0 = 0.05$ps at four different values of $\delta_3$. The blue curve corresponds to the analytical SY prediction [2].

In Fig. 1 the peak-power ratio between the lead (tallest) output soliton and the $N$-soliton input is 5.33 (black curve) for $\delta_3 = 0.01$, instead of 3.6 predicted by the SY formula (3). This result can be further improved by finding an optimal value of $\delta_3$, which in case of the 10-soliton is $\delta_3 = 0.0045$. The temporal and spectral dynamics in the optimal regime is demonstrated in the Fig.2 (a,b). We have performed an additional simulation, filtering the strongest soliton once it is separated from the parental $N$-soliton temporal slot at z ~ 0.3m and measured its propagation maintaining central wavelength to be upshifted to 948 nm (red vertical line in Fig. 2(b)). We have measured the full width at the half-maximum (FWHM) of the pulse for $\delta_3$ optimum to be 3.5fs, which corresponds very well to its peak power of 6.8kW as it should for fundamental NLS soliton. Compared with the injected pulse of $T_0 = 50$fs ($T_{FWHM} = 88$fs), the temporal width compression is estimated to be 25 times.

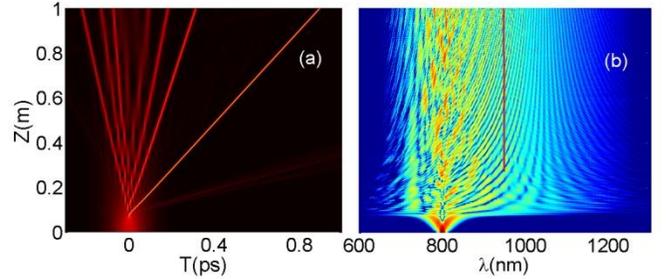

Fig. 2 (Color online) The generation of frequency-shifted fundamental solitons from the 10-soliton input with $T_0 = 0.05$ps at the optimal value of $\delta_3 = 0.0045$. Dynamics presented in (a) temporal and (b) spectral domains. The red vertical line represents the strongest soliton's evolution with the acquired constant wavelength of 948nm, once it is separated from the parental $N$-soliton temporal slot at z = 0.3m.

The optimal value of the TOD parameter $(\delta_3)_{opt}$ is only a function of the input soliton's N-number, which is shown in Fig. 3(a), while the so achieved peak-power ratio and the corresponding wavelength shift of the tallest soliton are depicted in Figs. 3(b) and 3(c), respectively. Figure 3(a) demonstrates that the optimal value of the TOD coefficient, $(\delta_3)_{opt}$, drops from 0.0220 to 0.0028, as the input soliton's order inc rises from $N = 3$ to 15. The number of distinct fundamental solitons produced by the fission the fission in the so the optimized regime is found to be $N-1$ for $3 < N < 8$ and $N-2$ for $9 < N < 15$, i.e., the loss in the number of the produced fundamental solitons is relatively small. Comparing the situations when $\beta_2$ and $\beta_3$ have opposite signs (i.e., $\delta_3 > 0$), or identical signs, $\delta_3 < 0$, we observed the mirror symmetry in the respective temporal-domain dynamics. On the contrary to the solitons' peak powers, the wavelength upshift and downshift significantly differ for fibers with opposite signs of $\delta_3$. The downshift of the lead-soliton's wavelength is limited by the proximity to the zero-dispersion point beyond which solitons cannot exist. For lowest-order $N$-soliton considered here, with $N = 3$, the wavelength shift is only ±18 nm added to the central wavelength, 800 nm, of the 3-soliton input. For larger $N$, the wavelength upshift up to 350 nm, with the power ratio of 8, is demonstrated by the solid red (upper) curve in Fig. 3(c). On the other hand, the downshift reaches only -164 nm for the 15-soliton, which is represented by the dashed blue (lower) curve in Fig. 3(c).

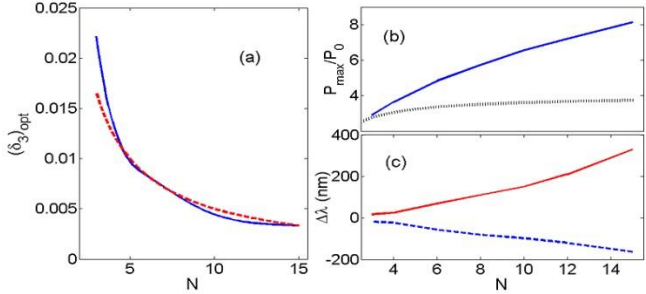

Fig. 3. (Color online) (a) The optimal value of $d_3 = \beta_3/(6\beta_2 T_0)$ as a function of the input-soliton's order (the blue solid curve). The red dashed curve shows the best fit to $\mathrm{const} \cdot N^{-1}$ (b), see the text. (b) The blue (upper) curve shows the ratio between the largest fundamental-soliton's peak power and the input peak power for both signs of $\delta_3$. The black dashed (lower) curve corresponds to the SY analytical result (3). The optimal values of $\delta_3$ is taken from panel (a). (c) The wavelength upshift and downshifts of the tallest generated fundamental soliton relative to the input wavelength of 800 nm. The upshift with $\delta_3 > 0$ and downshift with $\delta_3 < 0$ are represented, respectively, by the solid red and dashed blue (top and bottom) curves.

To better intuitively understand the power enhancement of the strongest ejected fundamental soliton, the dynamics of the 10-soliton was simulated with $\beta_3 = 0$ [Fig. 4(a)], and with gradually increasing $\beta_3$, see Figs. 4(b-c) and Fig. 2(a). The 10-soliton is injected into the fiber, which is followed by symmetric oscillations and creation of multiple peaks in Fig. 4(a). It is noteworthy, that in the most compressed state (Fig. 4(a), z~0.1) the 10-solitons peak power reaches 12.6kW and the FWHM is about 2.5 fs, but at the pedestal of this state appears a large "wing" structure of pulses tails that reaches the width of the original 50fs pulse. The action of the TOD, represented by $\beta_3$, gives rise to power and momentum transfer from one side of the pulse to the other in the course of quasi-elastic collisions, with taller peaks absorbing the energy from adjacent ones, and experiencing the acceleration (frequency shift), as seen in Figs. 4(b-c) and Fig. 2(a). These features of the TOD-mediated inter-soliton interactions were studied in detail earlier for soliton pairs [24-26], and more recently for the multi-soliton NC [10]. If the perturbation is weak, the "voracious" central peak captures the energy only from its closest neighbors, experiencing a conspicuous, but still not optimal, enhancement [Fig. 4(b)]. In this regime, the results are still close to those predicted by Eq. (3).

Further increase of $\delta_3$ enables the lead peak to absorb more energy and acquire a large frequency shift, to escape the temporal slot of the parent $N$-soliton, as seen in Fig. 4 (c) for $\delta_3 = 0.0025$. In this case, the largest peak power is 4.64 times higher than that of the injected 10-soliton. Finally, Fig. 2(a) demonstrates the 10-soliton's dynamics for the optimal value of $\delta_3 = 0.0045$, with the tallest soliton reaching the peak power 6.5 times higher than that of the parental 10-soliton. Further increase of $\delta_3$ causes the generation of strong dispersive waves, reducing the peak power and frequency shift of the lead soliton, as shown in Fig. 1 by the green and the orange curves.

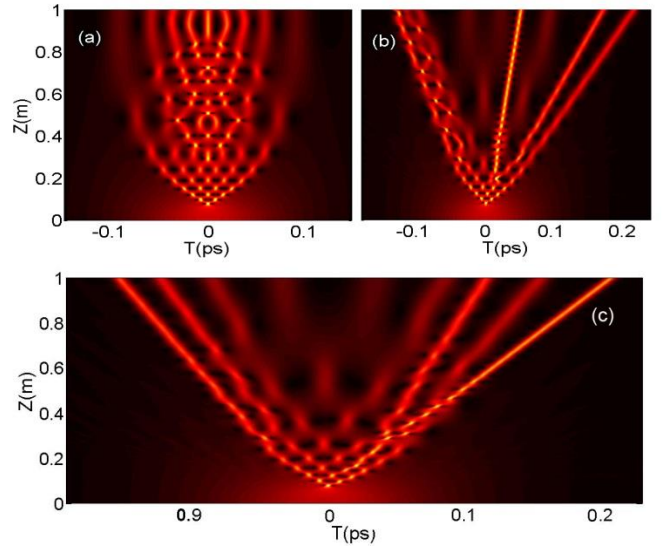

Fig. 4. (Color online) Simulations of the 10-soliton with $T_0 = 0.05$ps, using Eq. (1) with $s = 0$: (a) $\delta_3 = 0$; (b) $\delta_3 = 0.0012$; (c) $\delta_3 = 0.0025$.

A crude estimate of the optimum is possible if one notes that the compression of pulses under the action of the TOD is an adiabatic effect, i.e., a first-order one, in terms of the perturbation theory [1], while the degradation of the result due to the radiation loss is a second-order effect. Therefore, their contributions scale as $\delta_3 N$ and $(\delta_3 N)^2$, respectively, the optimum being attained at an equilibrium point, $\delta_3 N \sim (\delta_3 N)^2$, i.e., $\delta_3 N \sim$ const, which roughly agrees with the $\mathrm{const} \cdot N^{-1}$ fit to the optimum curve in Fig. 3(a).

If the shock term is restored in Eq. (1), the most effective energy capture and strongest soliton acceleration can be achieved even at lower values of $\delta_3$, as can be seen comparing Fig. 5(a) and Fig. 4(c), which pertain to the same value of $\delta_3 = 0.0025$. Figure 5(b) demonstrates another noteworthy effect occurring when the shock term is included: one of the fission products becomes a bound state of two solitons. Double-humped modes in the NLS equation with the TOD term were reported before as

"embedded solitons" [27], but they were unstable. In the present case, longer simulations demonstrate that the bound state is robust, but only if both the TOD and shock terms are present. These states (which are also known, as "soliton molecules", in other nonlinear-optics settings [28, 29]) will be considered in detail elsewhere.

The highest value of $N$ dealt with in this work is 15. For even higher orders of the input soliton, one can achieve still larger peak-power ratio and frequency shift, but the control over parameters of the soliton deteriorates due to interaction of the generated solitons with radiation [15-21]. Also noise driven dynamics (i.e. Modulation Instability) start to become important [5]. Besides that, the applicability of Eq. (1) to very strong deviations of the lead-soliton's wavelength from the wavelength of the $N$-soliton input may be limited.

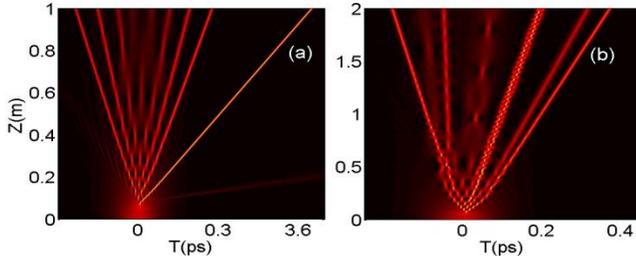

Fig. 4. (Color online) Simulations of the 10-soliton with $T_0 = 0.05$ps, using Eq. (1) with the shock term included (s=0.0085). (a) $\delta_3 = 0.0025$; (d) $\delta_3 = 0.0012$.

In conclusion, we have proposed and qualitatively explained the method of generating strongly compressed solitons from $N$-soliton inputs, with the tunable upshift and downshift of the central frequency, in Raman-inactive fibers. Such fibers allow for the fundamental solitons propagating with the acquired frequency, once ejected from the input N-soliton's temporal slot. The respective universal optimal value of the relative TOD strength has been found, depending solely on the order of the injected $N$-soliton. The optimal pulse-compression ratio significantly exceeds the prediction of the Satsuma-Yajima formula. The compression mechanism is valid in the presence of the self-steeping

The work of R. D. and B. A. M. is supported, in part, by the Binational (US-Israel) Science Foundation through grant no. 2010239